\documentclass[aps,pra,reprint,groupedaddress,showpacs]{revtex4-1} 

\usepackage{amsmath,amssymb,amsfonts}
\usepackage{graphicx}

\begin{document}


\title[Extended ray dynamics for integrable and chaotic optical microcavities]{Consequences of a wave-correction extended ray dynamics for integrable and chaotic optical microcavities}


\author{Pia Stockschl\"ader}
\author{Martina Hentschel}
\email{martina.hentschel@tu-ilmenau.de}
\affiliation{Institute for Physics, Technische Universit\"at Ilmenau - Weimarer Stra\ss{}e 25, 98693 Ilmenau, Germany}


\date{\today}

\begin{abstract}
Ray optics has proven to be an effcient and versatile tool to describe dielectric optical microcavities and their far-field emission based on the principle of ray-wave correspondence. 
Whereas often the well-known ray-optics at planar interfaces yields reasonable results, semiclassically and boundary-curvature induced corrections will become more important as the cavity size is further reduced. 
In this paper we summarize the various ray optics descriptions of optical microresonators, paying in addition special attention to the differences that arise between chaotic vs. non-chaotic (integrable or nearly integrable) resonator geometries, respectively. 
Whereas the far-field pattern in the chaotic case is known to be determined by the overlap of the unstable manifold with the leaky region, it results from the emission of trajectories with the smallest nonzero decay rates in the non-chaotic situation. 
We present an enhanced ray optical description, extended by wave-inspired (semiclassical) corrections, and discuss their consequences for the ray dynamics. 
In particular we find clear indications for the presence of attractors resulting from the non-Hamiltonian character of the extended
ray dynamics in phase space. 
We illustrate their impact on the far-field emission and show that it can considerably differ from the conventional ray description result. 
\end{abstract}


\maketitle


\section{Introduction}

Dielectric optical microcavities and microlasers have attracted much attention because of their possible applications in photonics and opto-electronics \cite{Vahala_microcavities,Chang_microcavities}.
In contrast to Fabry-Perot cavities, the light is confined by (total) reflection at the dielectric boundaries which reduces the space requirements as no mirrors are needed \cite{microcavities_review}.
Flat, quasi two-dimensional cavity geometries are especially interesting as they allow for easy in-plane integration in potential devices \cite{2d_microcavity}.
It has proven useful to describe these systems based on geometrical optics where the light propagation is approximated by rays \cite{chaotic_light_Stone,annular_billiard}.
In this efficient and easily implemented approach, the dielectric cavities can be considered as open billiards with the possibility of refractive escape and ray-splitting \cite{Blumel1996} being the origin of the openness.  
Using the concept of ray-wave correspondence, in analogy to the classical-quantum correspondence, much insight can be gained on the mode structure and the emission characteristics \cite{chaotic_light_Stone_nature,ray-wave-correspondence_chaotic,MH_husimi_epl}.

In this paper, we discuss and, whenever possible, summarize, important aspects concerning the ray description of dielectric optical microcavities. 
One focus will be on 
the differences between the description and understanding of cavities with predominantly chaotic and non-chaotic (integrable or nearly integrable) billiard dynamics, respectively. 
The differences in the spectra and the mode  structure of chaotic \textit{vs.}~integrable 
hard-wall billiards and open optical microcavities 
are well known and established both in experiments and electromagnetic wave simulations, see, \textit{e.g.}, \cite{quantum_chaos}. 
Here, however, we focus on the mechanisms that determine the far-field emission characteristics in chaotic \textit{vs.}~non-chaotic microcavities and -lasers from the point of view of a ray model description. 
Our objective of this theoretical study is to identify the mechanisms that determine the far-field characterisitcs, and their dependence on the relevance of wave-corrections, \textit{i.e.}, the size parameter of the cavity. 

Wave-inspired corrections to ray optics can become important for a reliable description especially of small cavities where the dimensions and the radius of the boundary curvature become comparable to the wavelentgh of the light \cite{Cao_wavelength-scale}. 
We discuss the influence of these correction terms for systems with curved and with planar boundaries, as well as for systems with chaotic and non-chaotic dynamics, respectively.
We illustrate our findings with several examples, namely, differently deformed disks and triangular cavities.

\section{Prediction of far-field emission: Chaotic \textit{vs.}~non-chaotic cavities}

We start with the comparison of chaotic and non-chaotic microcavities described with the normal (uncorrected or conventional) ray model. 
To determine the far-field emission from the ray model the dynamics of a large ensemble of rays is traced for a long time 
where time is measured in terms of the pathlengths that the trajectories have covered \cite{limacon}.
Whenever a ray trajectory hits the boundary it is specularly reflected and a part of the light, determined by Fresnel's law, can be transmitted.
The direction of the transmitted ray is given by Snell's law $\sin(\chi_\text{tr}) = n \sin(\chi_\text{in})$ where $\chi_\text{in}$ and $\chi_\text{tr}$ are the angle of incidence and transmission measured with respect to the boundary normal, respectively, and $n=n_1/n_2$ is the refractive index contrast between the cavity ($n_1$) and the surrounding medium ($n_2$).
The intensities of the reflected and the transmitted parts of the ray are given by the (Fresnel) reflection and transmission coefficients, $R$ and $T=1-R$ \cite{Jackson}.
When the light is totally internally reflected, \textit{i.e.} for $\sin(\chi_\text{in})>1/n$, no intensity is transmitted, \textit{i.e.} we have $R=1$ and $T=0$.
The far-field emission pattern is obtained by collecting the transmitted intensities of all trajectories in a given time interval as a function of the polar angle $\phi$. 

The ray dynamics in a two-dimensional billiard can be conveniently represented in a Poincar\'{e} surface of section of the full phase space spanned by the Birkhoff coordinates $(s,p)$ where $s$ is the position on the boundary and $p=\sin(\chi_\text{in})$ is the momentum component parallel to the boundary if the total momentum is normalized to $1$. 

\begin{figure*}
 \includegraphics[width=.98\textwidth]{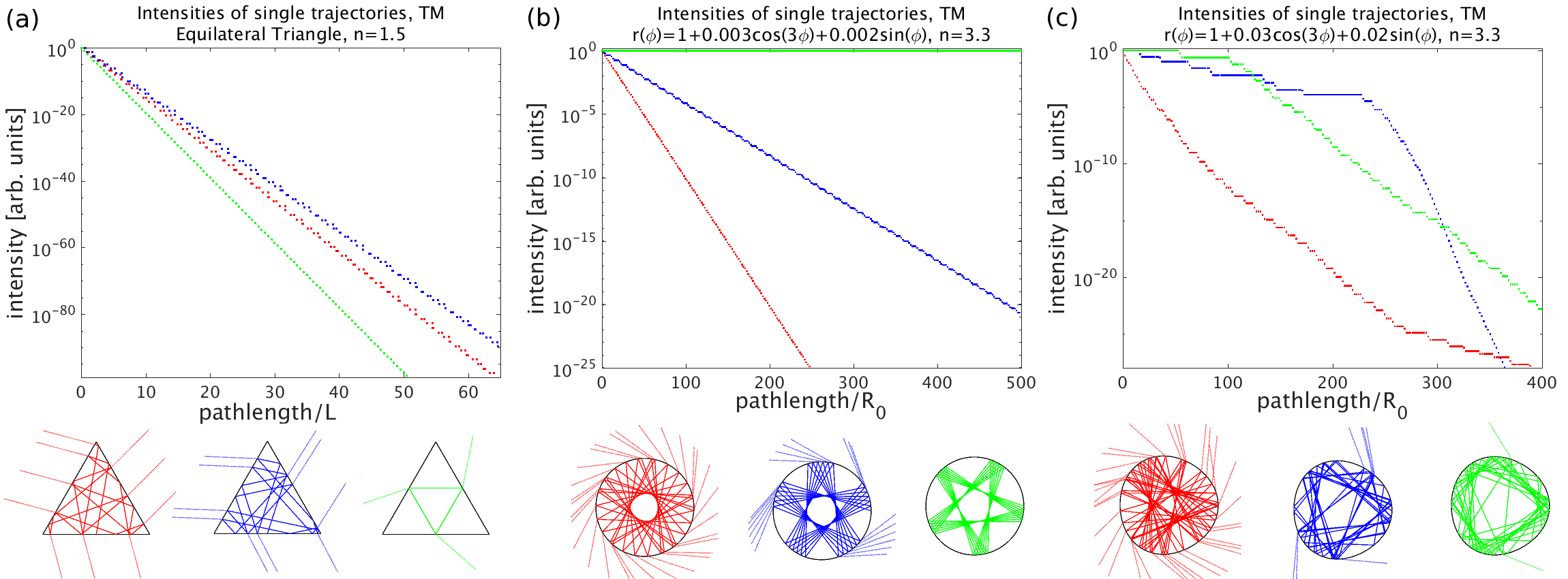}
 \caption{\label{fig:decay-rates_definition}
 Time evolution of the reflected intensity of exemplary trajectories 
 (a) in a cavity with $n=1.5$ in the shape of an equilateral triangle which has integrable 
  billiard dynamics for trajectories with starting conditions $s_0=1/6$ and $\chi_0=10^\circ$ (red), $\chi_0=18^\circ$ (blue), $\chi_0=30^\circ$ (green);
 (b) in a cavity with $n=3.3$ with a curved boundary and non-chaotic dynamics (boundary shape $r(\phi)=R_0(1+\epsilon_1\cos(3\phi)+\epsilon_2\sin(\phi))$ with $\epsilon_1=0.003$, $\epsilon_2=0.002$) for trajectories with starting conditions $\phi_0=0^\circ$ and $\chi_0=17^\circ$ (red), $\chi_0=18^\circ$ (blue), $\chi_0=19^\circ$ (green); 
 (c) in a cavity with a curved boundary and chaotic dynamics (boundary shape $r(\phi)=R_0(1+\epsilon_1\cos(3\phi)+\epsilon_2\sin(\phi))$ with $\epsilon_1=0.03$, $\epsilon_2=0.02$) for trajectories with starting conditions $\phi_0=0^\circ$ and $\chi_0=15^\circ$ (red), $\chi_0=30^\circ$ (blue), $\chi_0=45^\circ$ (green).
 The intensity is shown for TM polarized light. 
 For clarity, the trajectories are shown only for a couple of bounces not for the full time given in the intensity evolution plots. Note the difference in the intensity scales: The poor confinement in (a) is intimately related to the low refractive index of $n=1.5$ \textit{vs.}~$n=3.3$ used in (b) and (c).
 }
\end{figure*}

Now, we want to understand in more detail the mechanism that determines the far-field emission of a dielectric optical microcavity.
In the wave picture, the far-field is dominated by the emission of long-lived modes  of the cavity.
Analogously, in the ray picture, we expect that trajectories which are not immediatlely lost by refractive escape and rather keep `much' intensity inside the cavity 
will dominate the far-field.
In the case of chaotic cavities the set of these trajectories is well known and determined by the unstable manifold of the chaotic saddle
 \cite{deformed-cylinders_Schwefel,unstable_manifold1,unstable_manifold2}. Example trajectories are shown in Fig.~\ref{fig:decay-rates_definition}(c). 
Using the 
unstable manifold to predict the far-field emission of chaotic cavities is well established and has been successfully applied to many different geometries \cite{limacon,unstable_manifold1,unstable_manifold2,low-index,review_directional}.

In non-chaotic cavities, the mechanism for coupling out the light is different. 
For example, conservation of the angle of incidence implies either constant intensity or a continuous intensity drop along the trajectory. %
Again, we look for long-lived trajectories that we 
now, however, have to identify by their decay rates:
The emission pattern will be dominated by the emission of the trajectories with the smallest, yet nonzero decay rates. This is illustrated in Fig.~\ref{fig:decay-rates_definition}(a) for triangular, and in (b) for almost circular microcavities. 

In the case of regular, non-chaotic dynamics, the time evolution of the intensity of a single trajectory in an optical billiard is approximately given by
\begin{equation}
 I(\ell_\text{opt}) \sim I_0 \text{e}^{-\kappa \ell_\text{opt}}
 \label{eq:decay}
\end{equation}
where $\ell_\text{opt} = n \ell_\text{geo}$ is the optical pathlength that corresponds to the geometrical pathlength $\ell_\text{geo}$ that the trajectory has traveled inside the cavity with relative refractive index $n$. 
$\kappa$ is the decay rate of the considered trajectory (and amplification can be taken into account here, see \cite{PS_JOpt2016}).
The time is given by $t=\ell_\text{opt}/c$ with $c$ the speed of light in vacuum. 
A trajectory that is confined inside the cavity by total internal reflection for all times corresponds to $\kappa=0$, 
whereas, $\kappa>0$ indicates a trajectory with refractive losses. 
We have to keep in mind that the intensity is not a continuous function of time (or pathlength), rather, it can change only if the ray trajectory encounters the boundary, whereas it remains unchanged in between two reflections. 

To illustrate this behavior we show in Fig.~\ref{fig:decay-rates_definition} the evolution of the intensity of examplary trajectories in two different non-chaotic cavities and in a chaotic cavity.
In the non-chaotic case, Fig.~\ref{fig:decay-rates_definition}(a) and (b), we clearly see that the dependence of the intensity on the pathlength can be well approximated by a straight line in the semilogarithmic plot, indicating exponential decay.
In the first example, the equilateral triangle with relative refractive index $n=1.5$, Fig.~\ref{fig:decay-rates_definition}(a), the trajectories are poorly confined in the cavity. 
There exist no trajectories that are completely confined by total internal reflection \cite{PS_JOpt2016}, all trajectories lose their intensity rapidly, leading to high decay rates for all initial conditions, \textit{cf.}~Fig.~\ref{fig:decay-rates_definition}(a). Taking into account the amplification of light, such as of relevance in active, lasing sytems, can prove to be useful \cite{chaotic_explosions,chaotic_absorption,PS_JOpt2016} and helpful especially for a reliable prediction of the far-field patterns \cite{PS_JOpt2016}. 

The second example, an asymmetrically deformed disk with higher relative refractive index $n=3.3$, Fig.~\ref{fig:decay-rates_definition}(b), exhibits a broad distribution of decay rates. 
There are trajectories that are confined by total internal reflection for all times, \textit{i.e.}~not decaying at all, trajectories that exhibit incident angles above as well as below the critical angle, thus, decaying slowly, and quickly decaying trajectories that are not confined by total internal reflection.

This concept is, however, not applicable to optical billiards with chaotic dynamics.
In that case, the intensity decay is trajectory-specific as illustrated in 
Fig.~\ref{fig:decay-rates_definition}(c) for three different chaotic trajectories. 
A trajectory which starts above the critical line can stay there for many bounces such that the intensity does not change for some time. Eventually the trajectory reaches the leaky region and the intensity decreases (intermittency between leaky and trapped regions), and vice versa.
Consequently, its intensity 
cannot be approximated by an exponential decay as in (\ref{eq:decay}).

It is important to note that this argumentation is applicable only to the evolution of the intensity of single trajectories.
The total intensity inside a cavity, \textit{i.e.}~the accumulated intensity of all trajectories, decays exponentially in both cases, chaotic and non-chaotic, after some transition time \cite{chaotic_absorption,chaotic_explosions,fractal_weyl_microcavities}.

To summarize the first part, the far-field pattern of chaotic microcavities is determined by the unstable manifold (more precisely, the overlap of the unstable manifold of the chaotic saddle with the leaky region). For non-chaotic cavities, this well-known mechanism has to be modified, and it will be the trajectories with the smallest (non-vanishing) decay rates that determine the far-field emission. We point out that in both cases the far-field is constituted from the rays that \textit{(i)} can couple out from the cavity, and \textit{(ii)} survive in the long-time limit by minimizing the refractive intensity loss. 

\section{Extended ray optics: Conventional ray picture plus wave-inspired corrections} 

Despite the demonstrated success of the ray picture (see, \textit{e.g.}, \cite{deformed-cylinders_Schwefel,limacon,annular_billiard}), 
geometrical optics is only the zero-wavelength limit of electrodynamics. 
In the opitcal-microcavity reality, the light has a wavelength comparable to the system size.
Thus, deviations between the ray optics prediction and experiments or electromagnetic wave simulations are known to occur when the system size is on the order of several wavelengths \cite{microcavities_review,PS_JOpt2016,WiersigCao_ultrasmall}. 
To account for finite wavelength effects, wave-inspired correction terms can be included effectively in an extended 
ray optics description \cite{Fresnel_laws_curved,SchomerusHentschel_phase-space,PS_EPL2014,PS_JOpt2016}. 
These corrections have to be applied to 
the reflected and transmitted rays at each reflection point. A summary of all relevant effects and formulae 
is given in the Appendix, in the following we provide a brief qualitative description of the effects.

For the reflection of a light beam with finite width at a dielectric interface in (effectively) two dimensional geometries, deviations from the laws of geometrical optics are known \cite{Aiello_overview,TureciStone02,SchomerusHentschel_phase-space}:
A lateral shift of the reflected beam along the interface, the Goos-H\"anchen shift (GHS) \cite{GoosHaenchen,GoosHaenchen1949}, and a shift of the angles of reflection and transmission, called Fresnel filtering (FF) effect or angular Goos-H\"anchen shift \cite{TureciStone02,GoetteShinoharaHentschel2013}.
The effects act in different directions of the phase space, the Goos-H\"anchen shift acts in the $s$-direction (position on the boundary) and the Fresnel filtering effect acts in the $p$-direction (angle of incidence/momentum parallel to the interface) \cite{SchomerusHentschel_phase-space}.
As both effects are order-of-wavelength corrections having their origin in interference phenomena \cite{GoosHaenchen}, they can be interpretated in terms of a semiclassical correction to conventional ray optics. 
Note that these beam shifts break the law of reflection and Snell's law. 
Furthermore, the Fresnel filtering introduces non-Hamiltonian behavior to the ray dynamics, \textit{i.e.} a ray trajectory is no longer reversible in its time evolution \cite{nonHamiltonian}, and non-Hamiltonian features such as attractors and repellors 
appear as new structures in phase space. 
In many cases, an underlying structure of attractors is found to stabilize certain groups of orbits with slighty different intital conditions (that, in a chaotic cavity, would diverge in an uncorrected ray picture; see below, and, \textit{e.g.}, \cite{nonHamiltonian}).

Besides these two beam shifts effects, namely a) \textbf{the Goos-H\"anchen shift} and b) \textbf{the Fresnel filtering effect}, 
the extended ray picture requires in particular the use of c) \textbf{a beam-averaged 
 Fresnel reflection coeffcient}. The reason is, as for the Goos-H\"anchen shift and the Fresnel filtering, the finite extent of light beams in contrast to light rays. The light beam can be modelled as a bundle of light rays, comprised of individual light rays with, {\textit{e.g.},~a normal distribution. 
Each individual light ray will have its own Fresnel reflection coefficient depending on its angle of incidence, and depending on the presence or absence of a interface curvature \cite{Fresnel_laws_curved}. 
The main consequence of the beam-averaging is 
a delayed and smoothed-out onset of the regime of total internal reflection, \textit{i.e.}, a shift of the critical line to larger angles of incidence that is the more pronounced the higher the (local, convex) curvature is \cite{Fresnel_laws_curved,KotikHentschel}. 
We point out that this effect is, however,  present already at planar boundaries such as the triangular cavity. 
In particular, also the Brewster-angle feature (vanishing reflection of TE polarized light at the Brewster angle) is washed out, with crucial effects on the far-field emission as we shall see below. 

These three wave-inspired corrections introduce several different effects which can influence the far-field emission characteristics and explain the differences between the uncorrected and extended ray model far-fields: 
(\textit{i}) The change in the rules governing the ray propagation will change the system dynamics. 
Trajectories that dominate the far-field in the uncorrected case can lose their influence while other trajectories can gain influence, for example because they appear as attractors of the corrected, non-Hamiltonian ray dynamics. 
(\textit{ii}) The beam shifts in transmission can change the emission directions of the trajectories responsible for the far-field, both by the GH-shifting of the emission point along the curved interface, and by the FF-correction to the outgoing angle.
(\textit{iii}) 
The corrected reflection coefficients 
modify the critical line which will influence the emission pattern as well: 
There might be trajectories which are completely confined by total internal reflection in the uncorrected case but contribute to the far-field if the corrected reflection coefficients are applied. 
(\textit{iv}) As the semiclassical corrections incorporated into the extended ray model arise from beams (bundles of rays, to be described by averaged properties, cf.~Appendix), specific features in the Fresnel law such as the critical and the Brewster angle tend to be washed out by the averaging procedure. 
Consequently, the rather strong effect of the Brewster angle on the far-field emission of TE-polarized light (resulting in the characteristic TM-TE difference) 
will be reduced. 

\section{Far-field modifications due to extended ray optics}

We now apply the extended ray picture to several non-chaotic as well as chaotic examples in order to compare the appearence of their Poincar\'{e} surfaces of section and their far-field patterns in the conventional and enhanced ray optics, respectively. All details of the extended ray model, including formulae and parameters used here, are summarized in the Appendix below.

Firstly, we study two different triangles, the equilateral and the right isosceles triangle, as examples of systems with planar boundaries and integrable ray dynamics.
Secondly, we discuss asymmetrically deformed disks as examples of systems with curved boundaries which represent non-chaotic or chaotic ray dynamics depending on the choice of the deformation parameters. 

\subsection{Triangles as planar interfaces cavities} 
\label{sec:triangles}

Triangular cavities are examples for systems with planar interfaces only.
Like all polygons, triangles exhibit non-chaotic billiard dynamics \cite{billiards-in-polygons_1986,billiards-in-polygons_1996}.
Thus, we have to use the decay rates to identify those trajectories that will dominate the emission into the far-field.

Due to the unusual pseudointegrable dynamics of generic polygonal billiards the semiclassical treatment of polygons in the framework of ray-wave correspondence is difficult \cite{Wiersig_polygonal_billiards}.
However, it has been shown that ray-wave correspondence is fulfilled if the system is ``sufficiently open'' \cite{Wiersig_hexagon_PRA2003}.
That means, a ray-optical description can be applied to polygonal optical microcavities with a rather low refractive index.
We will discuss here triangular cavities with a relative refractive index $n=1.5$ that fulfill this requirement.

In an extended ray description of polygonal cavities only the angular shift (FF) is important for the determination of the far-field directions. 
The lateral shift along the (planar) boundary (GHS) amounts to a parallel shift of the reflected ray which does not alter the angles, and therefore not the emission directions of the trajectory.
The correction terms to the angle of reflection, $\Delta\chi^\text{ref}$, ranges from a fraction of a degree for angles of incidence smaller than the critical angle to a few degrees near the critical angle of incidence and vanishes quickly for supercritical incidence.
The correction to the angle of transmission, $\Delta\chi^\text{tr}$, amounts to several degrees in the vicinity of the critical angle. All details and functional dependences concerning the angular corrections can be found in the Appendix.

\begin{figure}
 \includegraphics[width=.48\textwidth]{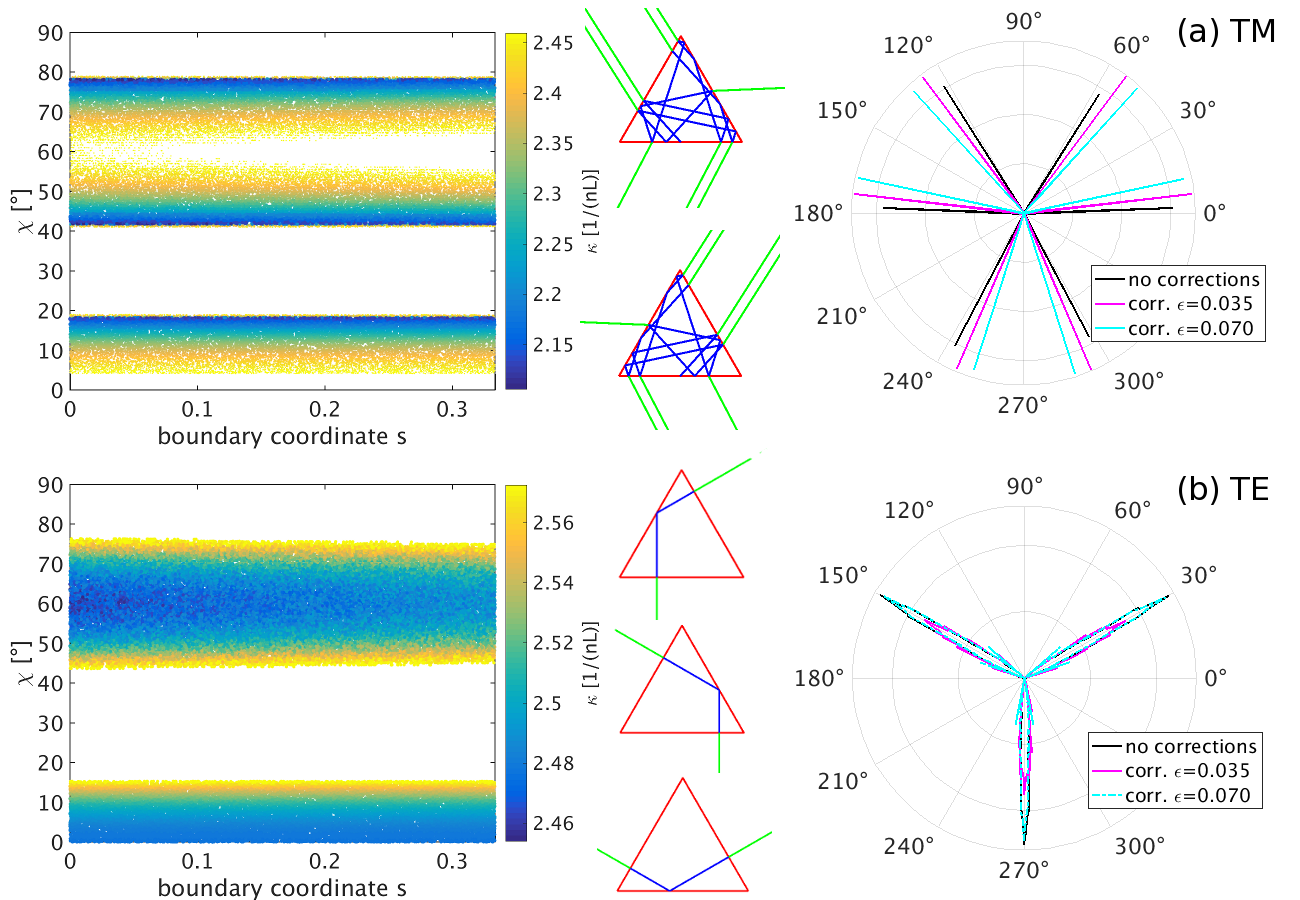} 
 \caption{\label{fig:equilateral_triangle}
 Results of the ray-optical description of the equilateral triangle with relative refractive index $n=1.5$ for both polarizations (a) TM and (b) TE. 
 \textit{Left:} Decay rates of the trajectories with initial conditions restricted to one of the three equivalent sides and positive angles of incidence.
 Only the smallest decay rates are shown.
 \textit{Center:} Examples of the trajectories with the smallest decay rates, trajectories with incident angles $18^\circ$, $42^\circ$ and $78^\circ$ for TM polarization, trajectories with incident angles $0^\circ$ and $60^\circ$ for TE polarization.
 \textit{Right:} Comparison of the far-field emission pattern calculated from the classical (black) and the corrected ray model with smaller (magenta) and larger (cyan) corrections.
 See the Appendix
 , especially section \ref{app:triangles}, for details concerning the corrections.
 }
\end{figure}

As a first example, we consider the equilateral triangle, see Fig.~\ref{fig:equilateral_triangle}.
For 
TE polarization, Fig.~\ref{fig:equilateral_triangle}(b), we identify the family of simple periodic trajectories with incident angles around $0^\circ$ and $60^\circ$ as the trajectories with the smallest nonzero decay rates, \textit{cf.}~the left panel of Fig.~\ref{fig:equilateral_triangle}(b). Shown is the  decay rate (color scale) as function of all possible initial conditions $(s, \chi)$.
Confirming our expectation, these trajectories indeed dominate the far-field emission: We find emission peaks perpendicular to all sides of the triangle.
This prediction agrees well with experiment and numerical simulations \cite{PS_JOpt2016,triangle_experiments}.
Including wave-inspired corrections to the ray model changes the far-field emission only in the details: 
As the incident angles of the dominant trajectories are far away from the critical angle, $\chi_c = \arcsin(1/n) \approx 41.8^\circ$, we do not expect the correction terms do strongly influence the 
far-field emission pattern. 
However, a closer look as in Fig.~\ref{fig:equilateral_TE_details} reveals some interesting details in form of extra peaks in the far-field, \textit{cf.}~\ref{fig:equilateral_TE_details}(c). Whereas the main emission directionality perpendicular to the sides is essentially kept, the non-Hamiltonian dynamics introduces attractors in phase space, see Fig.~\ref{fig:equilateral_TE_details}(b). As these will be a basin of attraction for trajectories, it is not surprising that extra peaks in the far-field, associated with these trajectories and attractors, may be observed, see Fig.~\ref{fig:equilateral_TE_details}(c) and sample trajectories in Fig.~\ref{fig:equilateral_TE_details}(d).

\begin{figure}
 \includegraphics[width=.48\textwidth]{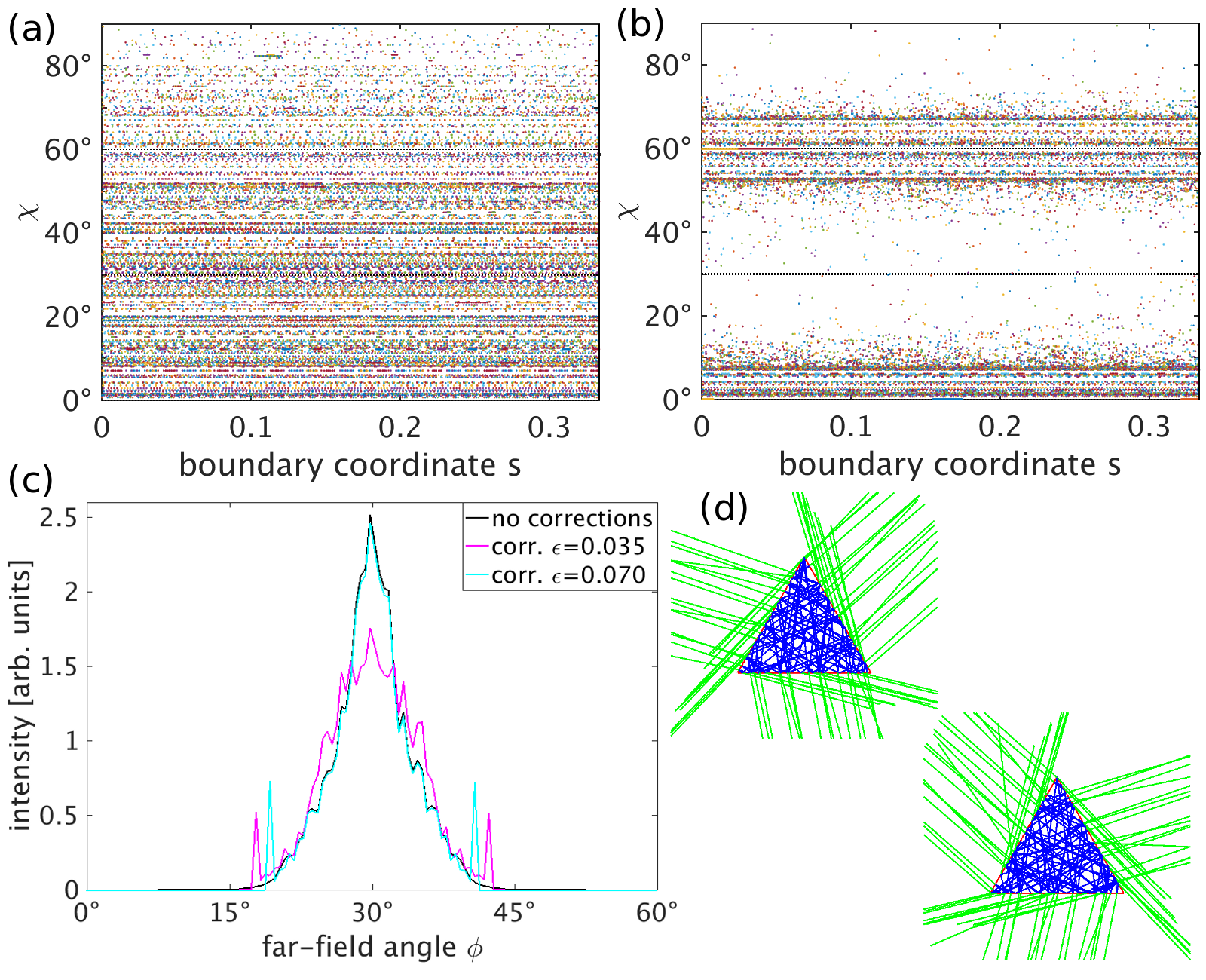} 
 \caption{\label{fig:equilateral_TE_details}
 Details of the ray-optical description of the equilateral triangle with relative refractive index $n=1.5$ for TE polarization.
 (a), (b) Poincar\'{e} surface of section of classical phase space of the normal and corrected ray dynamics with the larger corrections, respectively. 
 (c) Close-up of the for-field emission pattern shown in Fig.~\ref{fig:equilateral_triangle}(b) for incident angles $0^\circ\leq\chi\leq30^\circ$.
 (d) Examples of the trajectories pushed 
 towards the attractor in phase space by the beam shift corrections which are responsible for the extra peaks in the far-field.
 }
\end{figure}

For TM polarization, however, a different family of trajectories dominates the far-field, see Fig.~\ref{fig:equilateral_triangle}(a), as one can already deduce from the changes in the decay rate distribution shown in the left panel. 
In fact, a family of {\it non-periodic} trajectories with incident angles of approximately $18^\circ$, $42^\circ$, and $78^\circ$ has the smallest nonzero decay rates and determines the far-field (black lines in Fig.~\ref{fig:equilateral_triangle}(b)). The non-periodicity of the underlying trajectories is noteworthy, as often periodic orbits are known to govern the field of quantum chaos, at least for closed systems. 
The decay rate within this family of dominant trajectories increases fast away from the central orbit, in contrast to 
orbits relevant in the TE-case, and leads to the very sharp far-field paeks. 
As one of the incident angles 
is close to the critical incidence, $\chi_c\approx41.8^\circ$, the wave-inspired corrections are expected to have a strong effect on these trajectories and influence the far-field pattern (magenta and cyan curves in Fig.~\ref{fig:equilateral_triangle}(a)).
In a previous work, we have shown that the corrections are indeed necessary to obtain agreement between the ray model and wave simulations \cite{PS_JOpt2016}.

\begin{figure}
 \includegraphics[width=.48\textwidth]{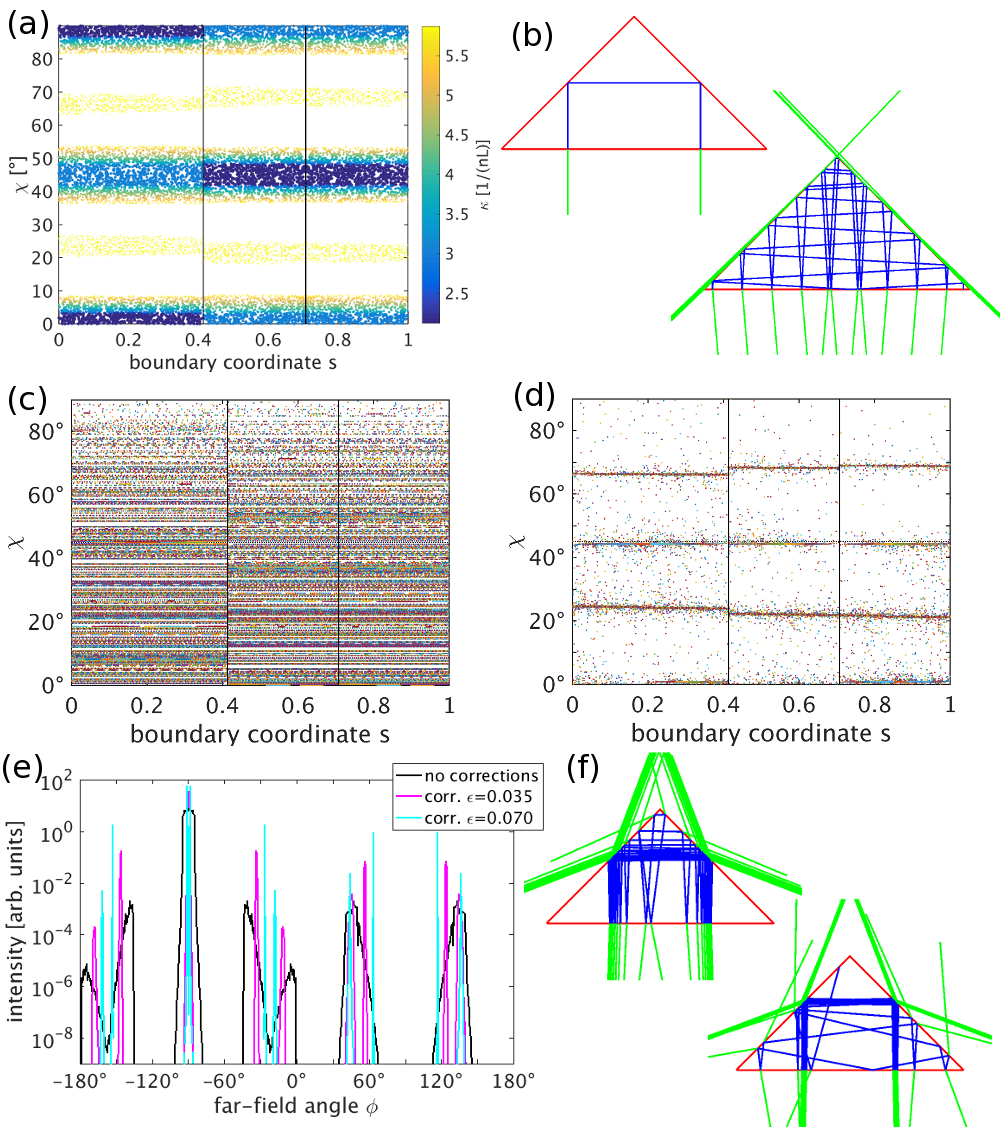} 
 \caption{\label{fig:right_isosceles_triangle}
 Results of the ray-optical description of the right isosceles triangle with relative refractive index $n=1.5$ for TE polarization. 
 (a) Decay rates of the trajectories with initial conditions restricted to positive angles of incidence. The vertical lines mark the positions of the corners along the boundary.
 (b) Simple periodic trajectory with incident angles $0^\circ$ on the hypotenuse and $45^\circ$ on the legs featuring the smallest decay rates (upper left). 
 Trajectory showing grazing emission with higher decay rate (lower right).
 (c) Poincar\'{e} section of the classsical phase space of the normal ray dynamics restricted to positive angles of incidence. 
 (d) Poincar\'{e} section of the corrected ray dynamics (for the larger, ``cyan'', correction).
 (e) Comparison of the far-field emission pattern calculated from the classical (black) and the corrected ray model with smaller (magenta) and larger (cyan) corrections. 
 Note the logarithmic scale used for the intensity.
 (f) Examples of the trajectories resulting from the corrected ray model which are stabilized on the simple periodic trajectoy shown in (b), upper left panel). 
 These trajectories show new emission directions from the legs resulting from the corrections to the reflection coefficient and are responsible for the changes in the far-field pattern.
 See the Appendix
 , especially section  \ref{app:triangles}, for details concerning the corrections.
 }
\end{figure}

As second example, we address the right isosceles triangle, see Fig.~\ref{fig:right_isosceles_triangle}.
Here, both polarizations behave very similar, because in both cases the relevant lowest-loss trajectory families live, in terms of phase space, sufficiently far away from the (sharp) Brewster angle feature. 
We will therefore restrict our discussion to the TE case.
The trajectories with the smallest nonzero decay rates, see Fig.~\ref{fig:right_isosceles_triangle}(b), are those with incident angles around $0^\circ$ on the hypotenuse and $45^\circ$ on the legs.
The complementary family of trajectories with incident angles around $45^\circ$ on the hypotenuse and $0^\circ$ on the legs has larger decay rates, but might still give some minor contribution to the far-field emission. It becomes visible on a logarithmic scale, see Fig.~\ref{fig:right_isosceles_triangle}(e), and the far-field emission directions of the right isosceles triangle are thus $-90^\circ$ (dominant), $45^\circ, 135^\circ$, and $-45^\circ, -135^\circ$ (from grazing/tangential emission).    

Including the angular corrections has an interesting effect in this example.
Due to the symmetry of the cavity and the closeness to the critical line of one angle of incidence of both simple periodic trajectories, the formation of attractors in phase space amounts to a stabilization of  trajectories with a broad distribution of initial conditions 
on these orbits, see the two examples shown in Fig.~\ref{fig:right_isosceles_triangle}(f). 
Whereas the far-field is still dominated by the emission peak into the $-90^\circ$-direction, there are now new contributions visible, see the cyan and magenta curves in Fig.~\ref{fig:right_isosceles_triangle}(e). 
They can be directly related to emissions from the stabilized orbits resulting both from non-grazing emission and near-critical, yet refractive, escape evident in the corresponding adjustment of the cyan and magenta emission peaks in Fig.~\ref{fig:right_isosceles_triangle}(e) in comparison to the classical ray picture denoted by the black curve.

This phenomenon, namely the stabilization of trajectory families when applying the Fresnel filtering correction, has been observed before \cite{nonHamiltonian,PS_JOpt2016} and can be related to the formation of attractors as the phase-space develops non-Hamiltonian characteristics as a result of the Fresnel-filtering correction that induces a Jacobian determinant different from 1 \cite{nonHamiltonian}. 
The depletion of certain phase-space regions, especially near the critical lines, in an extended ray dynamics due to the formation of attractors is 
clearly visible in Figs.~\ref{fig:equilateral_TE_details}(b) and \ref{fig:right_isosceles_triangle}(d), as well as in Figs.~\ref{fig:curved_nonchaotic}(b) and \ref{fig:curved_chaotic}(b) in the following section.

\subsection{Curved boundary cavities} 
\label{sec:curved}

In contrast to the planar case, the Goos-H\"anchen shift, in addition to the Fresnel filtering correction, has now to be included in the extended ray description, as a lateral shift along a curved boundary induces angular corrections as well. 
The shift along the interface, $D_\text{GH}$, is largest for incident angles around the critical angle where it is of the order of a few light wavelengths \cite{PS_EPL2014,SchomerusHentschel_phase-space}.
It vanishes for incident angles well below the critical angle; see 
the Appendix for more details. 
All wave corrections in the curved case are evaluated using the Fresnel laws derived for curved interfaces \cite{Fresnel_laws_curved,PS_EPL2014}.

We study asymmetrically deformed disks with a boundary given by 
\begin{equation}
 r(\phi)=R_0(1+\epsilon_1\cos(3\phi)+\epsilon_2\sin(\phi)) \:, 
 \label{eq:curved_boundary}
\end{equation}
using two different sets of deformation parameters $\epsilon_1$ and $\epsilon_2$ and a refractive index
$n=3.3$ typical for infrared light in GaAs-based devices.
For very small values of these parameters, the system shows regular ray dynamics, for larger deformation the system exhibits a mixed phase-space with a non-negligible chaotic component around the critical line.  
As for the right isosceles triangle, we restrict the discussion to TE polarization
in both examples. The results for TM polarization are comparable in general, except for Brewster angle related features when applicable. 

\subsubsection{Almost circular cavities.} 

For very small deformation parameters, $\epsilon_1=0.003$, $\epsilon_2=0.002$, the billiard 
shows mostly regular dynamics with a small chaotic component as seen in the Poincar\'{e} surface of section of the classical phase space shown in Fig.~\ref{fig:curved_nonchaotic}(a).
The trajectories with the smallest nonzero decay rates, see Fig.~\ref{fig:curved_nonchaotic}(c), are governed by a period-three stable orbit possessing islands above the critical line. 
They are well separated from other region of phase space by a gap that reflects the Brewster angle feature leading to increased refractive loss of corresponding trajectories in the TE case and, thus, large decay rates. 
They cross the critical line three times, so we expect tangential emission from the corresponding three points at the boundary. This yields six 
well-defined far-field emission peaks in the conventional ray picture, 
see Fig.~\ref{fig:curved_nonchaotic}(d). 
Note that these six peaks do not correspond to tangential emission from the points of highest boundary curvature, as one would expect in cavities without fast phase space diffusion as considered here \cite{chaotic_light_Stone}. 
Although the condition for total internal reflection tends to be violated more easily for higher curvature \cite{chaotic_light_Stone,chaotic_light_Stone_nature}, the cavity-specific phase space structure, dominated by three islands close to the critical line, alters the emission pattern. 

An important characteristics is the different height of these six emission peaks as two of them are strongly suppressed, and two are of intermediate height. This is due to the Brewster angle, $\chi_{B}=\arctan(1/n)$, where no TE polarized light is reflected. 
Figure \ref{fig:curved_nonchaotic}(e) shows a close-up of the decay rates in the vicinity of the Brewster and the critical angle (color scale as in Fig.~\ref{fig:curved_nonchaotic}(c), blue corresponding to low decay rates, yellow corresponding to higher decay rates).  
The decay rates of the individual trajectories are plotted at their starting points.
The 
asymmetry between the three emission regions along the boundary is clearly visible. 
The empty spots, corresponding to trajectories not contributing to the far-field, visible near $-90^\circ$ are directly related to the low peak heights into the far-field directions around $+150^\circ$ that would originate at the next reflection point. 
The small peak into $-30^\circ$ direction has its origin in a corresponding gap in the clockwise propagating sector (not shown). 
We point out that an effective reduction of the height (or, eventually, number) of far-field peaks in the TE-case, as a result of the presence of the Brewster angle, has been observed before, see \textit{e.g.}~\cite{limacon}. 

\begin{figure}
 \includegraphics[width=.48\textwidth]{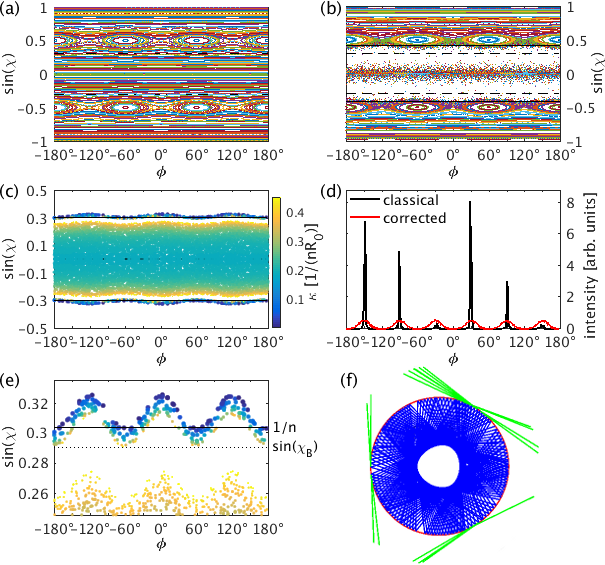} 
 \caption{\label{fig:curved_nonchaotic}
 Results of the ray-optical description of the asymmetrically deformed disk with non-chaotic dynamics given by boundary shape $r(\phi)=R_0(1+\epsilon_1\cos(3\phi)+\epsilon_2\sin(\phi))$ with $\epsilon_1=0.003$, $\epsilon_2=0.002$ with relative refractive index $n=3.3$ for TE polarization.
 Poincar\'{e} surface of section of the classical phase space (a) for conventional ray dynamics and (b) for extended ray dynamics.
 (c) Decay rates of the trajectories with initial conditions restricted to angles of incidence in the interval $|\sin(\chi)|\leq0.5$.
 Only the smallest nonzero decay rates are shown, for trajectories with incident angles with $|\sin(\chi)|>0.5$ all decay rates vanish. 
 (d) Comparison of the far-field emission pattern calculated from the classical (black) and the corrected (red) ray model.
See Appendix \ref{app:subsectioncurved} 
for details concerning the corrections.
 (e) Close-up of the decay rates around the critical angle (solid line) and the Brewster angle (dashed line).
 (f) Example of the family of trajectories with the smallest nonzero decay rates responsible for the far-field emission.
 }
\end{figure}

The wave-inspired correction terms introduce non-Hamiltonian features to the dynamics in an extended ray model, see Fig.~\ref{fig:curved_nonchaotic}(b). 
This becomes again manifest in the formation of repellors and attractors, 
in particular also in the depletion of the phase-space region around the critical angle. 
This significant change in the behavior of the system results in a considerable change in the far-field pattern, shown by the red curve in Fig.~\ref{fig:curved_nonchaotic}(d).
The peaks are now broadened, and are all of (nearly) the same height -- 
the Brewster angle is washed out when averaging over the reflection coefficients of the individual rays that consitute the light beam as the basis of the extended ray model.  
Consequently, the result resembles qualitatively the TM-case (not shown), and we conclude that the difference between TE- and TM-polarization is reduced in an extended \textit{vs.}~a conventional ray model.  

\subsubsection{Chaotic cavities.}

\begin{figure}
 \includegraphics[width=.48\textwidth]{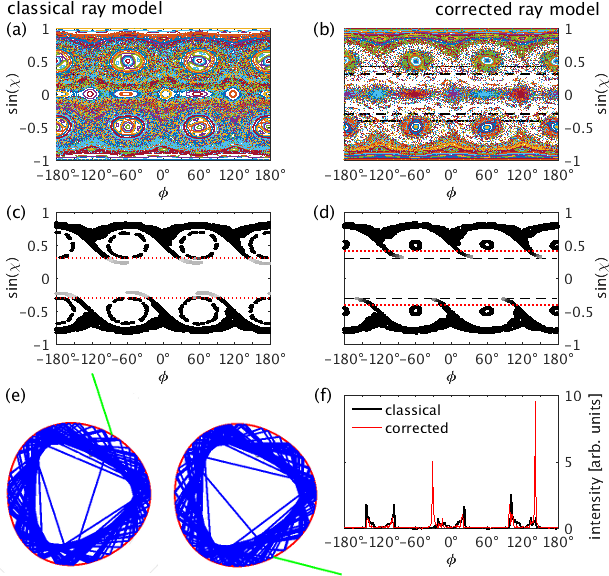} 
 \caption{\label{fig:curved_chaotic}
 As Fig.~\ref{fig:curved_nonchaotic}, but for 
 $\epsilon_1=0.03$, $\epsilon_2=0.02$. 
 (a), (b) Poincar\'{e} surface of section of classical phase space 
for the conventional and the extended ray model, 
 respectively. 
 (c), (d) Fresnel weighted unstable manifold of the chaotic saddle 
for the conventional and the extended ray model, 
 respectively. 
 (e) Examples of trajectories (conventional ray dynamics) starting on the unstable manifold that emit into the preferred far-field directions. 
 (f) Comparison of the far-field emission pattern calculated from the classical (black) and the corrected (red) ray model.
See Appendix \ref{app:subsectioncurved} 
for details concerning the corrections.
 }
\end{figure}

Choosing larger deformation parameters, here $\epsilon_1=0.03$ and $\epsilon_2=0.02$, the classical billiard dynamics of the system shown in Fig.~\ref{fig:curved_chaotic}(a) 
becomes partially chaotic
with a non-negligible chaotic component around the critical line.
In this case of a mixed phase space, we find the emission pattern (Fig.~\ref{fig:curved_chaotic}(f)) to follow the overlap of the unstable manifold of the chaotic saddle with the leaky region (Fig.~\ref{fig:curved_chaotic}(c)), as is known from the (fully) chaotic case \cite{unstable_manifold1,limacon}. 
The emission from trajectory representatives of the unstable manifold is illustrated in Fig.~\ref{fig:curved_chaotic}(e). 

This examples illustrates how stable islands near the critical line determine the far-field emission characteristics. 
The presence of three islands channels the unstable manifold such that the far-field has a three-peak structure as well (peaks centered around $0^\circ$ and $\pm 120^\circ$), \textit{cf.}~Fig.~\ref{fig:curved_chaotic}(f), that are, however, not exactly equivalent due to the cavity asymmetry. 
Furthermore, each of the three peaks has a pronounced dip in its center that reflects the fact that the stable island reaches into the leaky region formed in between the critical lines. 
Indeed, assuming a higher refractive index ($n \geq 4$) such that the islands now lie outside the critical lines, removes the dip and leads to a central maximum in each of the three emission directions. 

Introducing the wave-inspired corrections to the ray-model has, in principle, the same effects on the dynamics of the system as in the almost circular 
case, see Fig.~\ref{fig:curved_chaotic}(d), most notably an increase of the chaotic phase-space region and the development of attractors and repellors as signatures of the non-Hamiltonian dynamics. Figure \ref{fig:curved_chaotic}(b) illustrates the formation of attractors and repellors particularly nicely. 
Especially evident is a shift of the critical line to higher incident angles as compared to the the conventional ray model, see Fig.~\ref{fig:curved_chaotic}(d), \textit{i.e.}, the leaky region increases. 
The intricate interplay of this effect, the modification of the unstable manifold in the extended ray picture, and the beam-averaging of the Fresnel reflection coefficient amounts to a noticable change in the intensity of the far-field emission directions, although these themselves remain essentially the same. 
The latter can be attributed to the robustness of far-field emission based on the unstable manifold. 
However, our simulations suggest that effects of an extended ray dynamics can lead to noticable 
changes of the far-field characteristcs. 

\section{Discussion and conclusion}

In summary, we have discussed a ray-optics based prediction 
of the far-field emission properties of dielectric optical microcavities paying special attention to differences between chaotic and non-chaotic micrcavities and features that are relevant, for example, in ultrasmall cavities when wave-effects require the use of an extended ray picture. 
On the one hand side, we studied the differences in the description of systems with chaotic and non-chaotic (mixed and integrable) classical ray dynamics.
Whereas it is well known that the far-field emission of chaotic cavities is determined by the unstable manifold of the chaotic saddle, we argue here that, in the case of non-chaotic cavities, the emission is dominated by the trajectories with the smallest nonzero decay rates.
On the other hand side, we included wave-inspired corrections to the ray-model which are known to be important to obtain agreement between the prediction from the extended ray model with experiments and/or wave simulations. 
The extended ray model takes a) the Goos-H\"anchen shift, b) the Fresnel filtering correction (in reflection and transmission) as well as corrections to the Fresnel reflection coefficient for planar and for curved cavity boundaries into account. 
An c) averaged Fresnel reflection coefficient has to be applied in order to take all wave-inspired corrections, that ultimately result from using light beams rather than individual light rays, consistently into account. 

Our main findings in the context of this extended ray model are: 
\textit{(i)} 
If the trajectories that dominate the far-field emission pattern (identified by the unstable manifold or via the decay rates) are (not) strongly influenced by these 
corrections, then the far-field emission is (not) affected by the inclusion of the correction terms and will change in comparison to the conventional ray picture (that will thus be not reliable in this case, see, \textit{e.g.}, \cite{PS_JOpt2016} for the equilateral triangle case). 
Moreover, 
\textit{(ii)} the extended ray optics may change the phase-space structure so substantially  (introducing non-Hamiltonian features, shifting the critical line) that far-field emission directions and/or intensities of conventional and extended ray picture do not coincide any more. 
This happens in particular when the phase-space structures that determine the far-field emission pattern (the overlap of the unstable manifold with the leaky region  in the chaotic case, or the structures representing the trajectories with the smallest nonzero decay rates otherwise) are affected by these corrections (\textit{i.e.}, the formation of attractors in the vicinity of the critical line or the formation of repellors for smaller $|\sin\chi|$). 
In all theses cases, the extended ray picture can give more reliable results than the conventional ray picture. 

\appendix*
\section{The extended ray model}
\label{app:wave-corrections}

The evaluation of the wave-inspired corrections to the ray model used here is based on an expectation value approach \cite{Porras_moment-method,PS_EPL2014}. As a curvature of the microcavity is of special importance, we distinguish between planar and curved interfaces and give an outline of both cases in the following. 

\subsection{Wave-inspired corrections at planar interfaces}

In the planar case, the beam is expanded in plane waves. 
The calculations are restricted to the interface, denoted by the $z$-axis.
The angles of incidence $\chi$ are related to the direction of the incident plane wave via $k_z^\text{in}=nk\sin(\chi)=nkp$ with the wavenumber $k=2\pi/\lambda$ where $\lambda$ is the vacuum wavelength of the light beam and $p=\sin(\chi)$.
The incident beam is given by 
\begin{equation}
 E_I(z) = \int_0^1\! \text{d} p \, e_I(p)\text{e}^{\text{i}nkpz} ,
 \label{eq:inc_beam_planar}
\end{equation}
where the transverse beam profile $e_I(p)$ is supposed to be a narrow distribution in momentum $p$ with expectation value $p=p_0$ corresponding to the central angle of incidence $\chi_\text{in}$ with $p_0=\sin(\chi_\text{in})$.

The reflected beam is obtained by applying the Fresnel reflection coefficients to the incident beam profile, $e_R(p)=\rho(p) e_I(p)$.
The Fresnel equations \cite{Jackson} expressed in $p=\sin(\chi)$ read for both, TM and TE, polarizations
\begin{equation}
 \begin{aligned}
  \rho_\text{TM} (p) &= \frac{n \sqrt{1 - p^2} - \sqrt{1 - n^2 p^2}}{n \sqrt{1 - p^2} + \sqrt{1 - n^2 p^2}}\\
  \rho_\text{TE} (p) &= \frac{\sqrt{1 - p^2} - n \sqrt{1 - n^2 p^2}}{\sqrt{1 - p^2} + n \sqrt{1 - n^2 p^2}}
 \end{aligned}
 \label{eq:Fresnel}
\end{equation}
where TE polarization shows the Brewster angle $\chi_B$, given by $\tan(\chi_B)=1/n$ with $\rho_\text{TE} (\sin(\chi_B))=0 $.
With that, the reflected beam in real space is 
\begin{equation}
 E_R(z) = \int_0^1 \! \text{d} p \, e_R(p)\text{e}^{\text{i}nkpz} .
 \label{eq:ref_beam_planar}
\end{equation}

We define the position of incidence of the beam on the interface as the expectation value of $z$ of the incident beam profile
\begin{equation}
 \left\langle z \right\rangle_\text{in}
 = \frac{\int_{-\infty}^\infty \!\text{d} z \, E^*_I(z)\, z\, E_I(z)}{\int_{-\infty}^\infty \!\text{d} z \left| E_I(z) \right|^2} .
 \label{eq:inc_pos_planar}
\end{equation}
The (mean) position of the reflected beam is analogously given by the $z$-expectation value of the reflected beam profile
\begin{equation}
 \left\langle z \right\rangle_\text{ref}
 = \frac{\int_{-\infty}^\infty \!\text{d} z \, E^*_R(z)\, z\, E_R(z)}{\int_{-\infty}^\infty \!\text{d} z \left| E_R(z) \right|^2} .
 \label{eq:ref_pos_planar}
\end{equation}
The difference between these two positions is the Goos-H\"anchen shift along the boundary
\begin{equation}
 D_\text{GH} = \left\langle z \right\rangle_\text{ref} - \left\langle z \right\rangle_\text{in}.
 \label{eq:GHS_planar}
\end{equation}
If the incident beam profile in angular momentum space is symmetric with respect to the central momentum component $p_0$, Eq.~\eqref{eq:inc_pos_planar} yields $\left\langle z \right\rangle_\text{in} = 0$ and the position of incidence marks the origin of the chosen coordinate system.
Hence, the Goos-H\"anchen shift becomes $D_\text{GH} = \left\langle z \right\rangle_\text{ref}$ which is the case in all the examples that we consider here.

The expectation values of the momentum $p$ calculated with the angular profiles give the directions of the beams.
The mean angle of incidence $\chi_\text{in}$ corresponds to
\begin{equation}
 \left\langle p \right\rangle_\text{in} 
 = \frac{\int_0^1\! \text{d} p \, e^*_I(p)\, p\, e^{}_I(p)}{\int_0^1\! \text{d} p \left| e_I(p) \right|^2}.
 \label{eq:inc_direction_planar}
\end{equation}
By the choice of the incident angular profile it is $\left\langle p \right\rangle_\text{in} = p_0 = \sin(\chi_\text{in})$.
The mean angle of reflection $\chi_\text{ref}$ can be calculated from the $p$-expectation value of the reflected angular profile
\begin{equation}
 \begin{aligned}
  \left\langle p \right\rangle_\text{ref} 
   = \frac{\int_0^1\! \text{d} p \, e^*_R(p)\, p\, e^{}_R(p)}{\int_0^1\! \text{d} p \left| e_R(p) \right|^2}\\
   = \frac{\int_0^1\! \text{d} p \, p\, R(p)\left| e_I(p) \right|^2}{\int_0^1\! \text{d} p \, R(p)\left| e_I(p) \right|^2}
 \end{aligned}
 \label{eq:FF_planar}
\end{equation}
with $\left\langle p \right\rangle_\text{ref} = \sin(\chi_\text{ref})$.
Here, we have introduced the intensity reflection coefficient $R(p) = \left| \rho(p) \right|^2$.
The difference between the mean angle of incidence and of reflection is the angular shift constituting the Fresnel filtering effect, $\Delta\chi_\text{FF} = \chi_\text{ref} - \chi_\text{in}$.

Analogously to the definition of the mean direction of the reflected beam in Eq.~\eqref{eq:FF_planar}, we can define the mean direction of the transmitted beam as the $p$-expectation value of the transmitted angular profile
\begin{equation}
  \left\langle p \right\rangle_\text{trans} 
   = \frac{\int_0^{1/n}\! \text{d} p \, p\, T(p)\left| e_I(p) \right|^2}{\int_0^{1/n}\! \text{d} p \, T(p)\left| e_I(p) \right|^2}
 \label{eq:trans_direction_planar}
\end{equation}
with the intensity transmission coefficient $T(p) = 1 - R(p)$.
Here, the integration runs only over those momenta $p\leq1/n$ for which partial transmission is possible.
The mean angle of transmission $\chi_\text{trans}$ is given by $n \left\langle p \right\rangle_\text{trans} = \sin(\chi_\text{trans})$ where the change in direction between the incident and the transmitted ray due to the refractive index contrast is taken into account.
The difference between the mean angle of transmission calculated from the transmitted beam and the angle of transmission expected from Snell's law is the Fresnel filtering effect in transmission
\begin{equation}
 \Delta\chi_\text{FF}^\text{trans} = \chi_\text{trans} - \arcsin(n\sin(\chi_\text{in} )).
 \label{eq:FF_planar_trans}
\end{equation}
A lateral shift along the interface, in analogy to the Goos-H\"anchen shift, does not occur in transmission \cite{GoetteShinoharaHentschel2013}.
As there is no phase shift between the incident and the transmitted waves \cite{Jackson} there is no spatial shift of the transmitted beam.

Due to the different weight given by the reflection coefficients of the components of the incident beam the intensity of the reflected beam can deviate from the intensity of the reflection of the central component of the incident beam.
The intensity of the reflected beam is simply given by the expectation value of the intensity reflection coefficient
\begin{equation}
 \begin{aligned}
  \left\langle I \right\rangle_\text{ref} 
   = \frac{\int_0^1\! \text{d} p \, e^*_R(p) e^{}_R(p)}{\int_0^1\! \text{d} p \left| e_I(p) \right|^2}\\
   = \frac{\int_0^1\! \text{d} p \, R(p)\left| e_I(p) \right|^2}{\int_0^1\! \text{d} p \, \left| e_I(p) \right|^2}.
 \end{aligned}
 \label{eq:I_ref_planar}
\end{equation}
As we assume no absorption or any other losses, $\left\langle I \right\rangle_\text{trans} = 1 -\left\langle I \right\rangle_\text{ref}$ is still valid.

\begin{figure}
 {\centering
  \includegraphics[width=.48\textwidth]{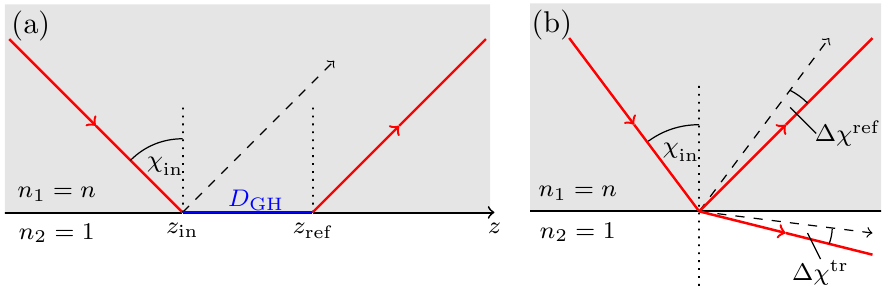}
  
 }
 \caption{\label{fig:rays_at_interface}
 Overview of the extended ray model.
 The reflected ray 
 experiences corrections to the position, the angle of reflection and the reflected intensity.
 The light is transmitted with a corrected intensity and a correction to Snell's law.
 }
\end{figure}

To summarize, the extended ray model provides an effective, beam-based ray description. 
See Fig.~\ref{fig:rays_at_interface} for a schematic overview of the wave-inspired corrections.
They consist in \textit{(i)} the Goos-H\"anchen correction that affects the position of the reflected beam along the interface boundary,
\begin{equation}
 s_\text{ref}(\chi_\text{in}) = s_\text{in} + D_\text{GH}(\chi_\text{in}),
 \label{eq:Goos-Haenchen_app}
\end{equation}
\textit{(ii)} the Fresnel filtering correction that corrects the reflection as well as Snell's law,
\begin{equation}
 \chi_\text{ref} (\chi_\text{in}) = \chi_\text{in} + \Delta\chi^\text{ref}(\chi_\text{in}),
 \label{eq:angle_ref_app}
\end{equation}
\begin{equation}
 \chi_\text{tr}(\chi_\text{in}) = \arcsin(n\sin(\chi_\text{in})) + \Delta\chi^\text{tr}(\chi_\text{in}),
 \label{eq:angle_trans_app}
\end{equation}
and (iii) in a correction of Fresnel's law for the reflection and transmission coefficients resulting from an averaging over the individual rays that constitute the light beam. 

\subsection{The correction terms used for the triangular cavities}
\label{app:triangles}

In the case of triangular cavities discussed in the body of the paper and for polygonal cavities in general the Goos-H\"anchen shift does not play a role for the determination of the far-field emission in the ray model.
The lateral shift of the reflected ray along the interface leads only to a parallel shift of the whole trajectory which does not affect the far-field emission direction.
Therefore, we only need to consider the angular corrections in reflection and transmission and the corrected intensities.

\begin{figure}
 \includegraphics[width=\linewidth]{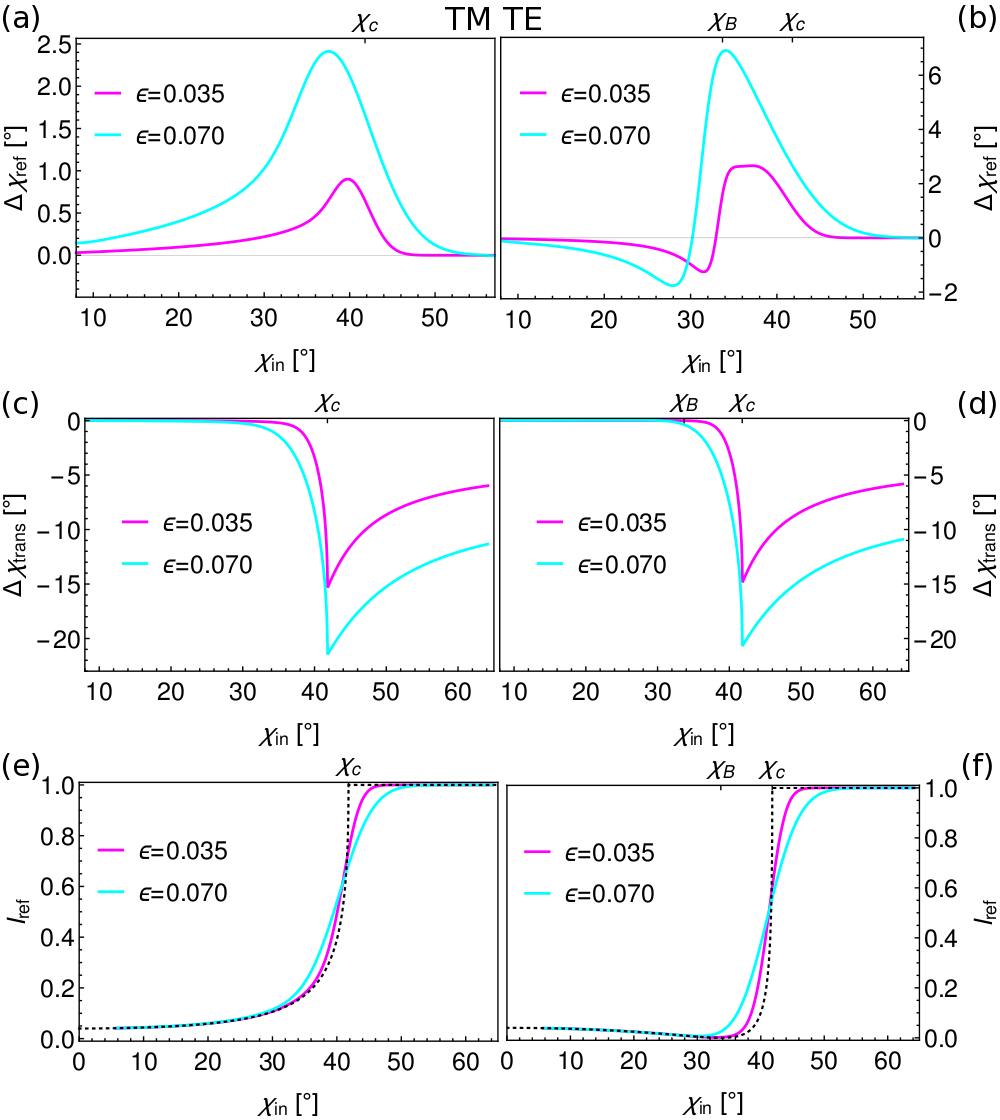}
 \caption{\label{fig:corrections_triangles}
 Wave-inspired corrections at a planar interface with relative refractive index $n=1.5$ for both polarizations, TM and TE shown in left and right column, respectively, and two different momentum distributions, $\epsilon=0.035$ (magenta) and $\epsilon=0.070$ (cyan), according to Eq.~\eqref{eq:gaussian_planar}.
 }
\end{figure}

For an evaluation of the formulas for the correction terms at planar interfaces, we chose a Gaussian with mean value $p_0$ and standard deviation $\epsilon$ as the transverse beam profile in momentum space,
\begin{equation}
 e_I(p) = \frac{1}{\sqrt{2\pi}\epsilon}\text{e}^{-\frac{(p-p_0)^2}{2\epsilon^2}}.
 \label{eq:gaussian_planar}
\end{equation}
In Fig.~\ref{fig:corrections_triangles} we show the resulting corrections for a beam incident at a dielectric interface with relative refractive index $n=1.5$ and for two different momentum distributions with $\epsilon=0.035$ and $\epsilon=0.070$, respectively.  
We use these results for the enhanced ray optics description of the triangular cavities discussed in Section \ref{sec:triangles}.

\subsection{Wave-inspired corrections at curved interfaces}
\label{app:subsectioncurved}

Analogously to the expectation value approach at planar interfaces, the beam shifts at convexly curved interfaces can be defined as expectation values.
We assume circular symmetry and make use of the corrected Fresnel coefficients for curved interfaces provided in Ref.~\cite{Fresnel_laws_curved}.
The geometry and the notations used in this section are clarified in Fig.~\ref{fig:rays_in_circle}.
Due to the radial symmetry, polar coordinates $(r,\alpha)$ are used and cylinder functions, Bessel and Hankel functions, are the appropriate basis functions.
Angular momentum conservation leads to a relation between the angular wavenumber $m$ of the cylinder function $J_m$ and the angle of incidence $\chi$ \cite{Fresnel_laws_curved,chaotic_light_Stone}
\begin{equation}
 \sin(\chi) = \frac{m}{nkR}.
 \label{eq:angle_curved}
\end{equation}
According to Ref.~\cite{Fresnel_laws_curved}, 
the reflection coefficients at a convexly curved interface read
\begin{equation}
 \rho_c = \frac{\cos(\chi) + i\mathcal{F}}{\cos(\chi) - i\mathcal{F}}
 \label{eq:Fresnel_corrected}
\end{equation}
with 
\begin{equation*}
 \mathcal{F}^\textrm{TE} = n\frac{H^{(1)'}_m(kR)}{H_m^{(1)}(kR)} 
 \quad \text{and} \quad 
 \mathcal{F}^\textrm{TM} = \frac{1}{n^2}\mathcal{F}^\textrm{TE} 
\end{equation*}
where $H_m^{(1)}$ is the Hankel function of the first kind and prime denotes the derivative with respect to the full argument.

\begin{figure}
 \includegraphics[width=.48\textwidth]{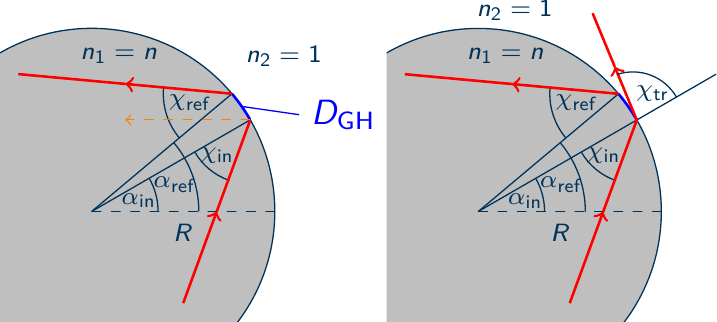}
 \caption{\label{fig:rays_in_circle}
 Beam shift effects at convexly curved interfaces. Left: Although the Goos-H\"anchen shift takes place along the interface as before in the planar case, light rays reflected with and without (orange dashed) the Goos-H\"anchen displacement do not propagate parallel any more. In addition to the lateral shift, the Fresnel filtering correction implies that $\chi_\text{ref} > \chi_\text{in}$ around critical incidence and influences the angle of transmittance $\chi_\text{tr}$ as well, see right panel.
 }
\end{figure}

The beams are conveniently expanded in polar coordinates $(r,\alpha)$ using Bessel functions \cite{SchomerusHentschel_phase-space}.
The incident light beam at the interface $r=R$, then, is 
\begin{equation}
 E_I(\alpha) = \sum_m e_I(m) \text{e}^{\text{i}m\alpha} J_m(nkR).
 \label{eq:inc_beam_curved}
\end{equation} 
The transverse beam profile in angular momentum space $e_I(m)$ is chosen to be a narrow distribution in $m$ with expectation value at the central angular wavenumber $m_0$ corresponding to $\chi_\text{in}$ using Eq.~\eqref{eq:angle_curved}, $\sin(\chi_\text{in})=m_0/(nkR)$.
The reflected beam is obtained by applying the corrected Fresnel reflection coefficients for convexly curved interfaces given in Eq.~\eqref{eq:Fresnel_corrected} to the incident beam profile, $e_R(m)=\rho_c(m) e_I(m)$, giving 
\begin{equation}
 E_R(\alpha) = \sum_m e_R(m) \text{e}^{\text{i}m\alpha} J_m(nkR).
 \label{eq:ref_beam_curved}
\end{equation}

The beam shifts are obtained from the expectation values of the polar angle and the angular wavenumber \cite{PS_EPL2014}.
The mean position of incidence on the boundary is given by the expectation value of the polar angle $\alpha$ with respect to the incident beam
\begin{equation}
 \langle \alpha \rangle_\text{in} 
 = \frac{\int_{-\pi}^{\pi}\! \text{d} \alpha \, E_I^*(\alpha)\, \alpha\, E_I(\alpha)}{\int_{-\pi}^{\pi}\! \text{d} \alpha \, E_I^*(\alpha) E_I(\alpha)}. 
 \label{eq:inc_pos_curved}
\end{equation}
Correspondingly, the mean position of reflection of the beam is given by the $\alpha$-expectation value of the reflected beam profile
\begin{equation}
 \langle \alpha \rangle_\text{ref} 
 = \frac{\int_{-\pi}^{\pi}\! \text{d} \alpha \, E_R^*(\alpha)\, \alpha\, E_R(\alpha)}{\int_{-\pi}^{\pi}\! \text{d} \alpha \, E_R^*(\alpha) E_R(\alpha)}. 
 \label{eq:ref_pos_curved}
\end{equation}
The lateral shift $D_\text{GH}$ along the interface, given in multiples of the vacuum wavelength $\lambda$, is obtained from the possible difference between the mean positions of reflection and incidence
\begin{equation}
 \frac{D_\text{GH}}{\lambda} = \frac{nkR}{2\pi} \left( \langle \alpha \rangle_\text{ref} - \langle \alpha \rangle_\text{in} \right).
 \label{eq:GHS_curved}
\end{equation}
If the incident beam profile is symmetric Eq.~\eqref{eq:inc_pos_curved} yields $\left\langle \alpha \right\rangle_\text{in}=0$ and the position of incidence marks the origin of the polar angle.
Hence, the expression for the Goos-H\"anchen shift $D_\text{GH}$ simplifies accordingly.

We can obtain the mean angles of incidence and reflection from the expectation values of the angular wavenumber.
The expectation value of $m$ with respect to the incident angular profile
\begin{equation}
 \langle m \rangle_\text{in} = \frac{\sum_m e_I^*(m)\, m\, e_I (m)}{\sum_m e_I^*(m)e_I(m)}
 \label{eq:inc_direction_curved}
\end{equation}
equals the mean angular wavenumber, $\langle m \rangle_\text{in} = m_0$, corresponding to the chosen angle of incidence $\chi_\text{in}$ with $\chi_\text{in} = m_0/(nkR)$.
The angle of reflection $\chi_\text{ref}$ is obtained from the $m$-expectation value of the reflected angular profile
\begin{equation}
\begin{aligned}
 \langle m \rangle_\text{ref} &= \frac{\sum_m e_R^*(m)\, m\, e_R (m)}{\sum_m e_R^*(m)e_R(m)} \\
 &= \frac{\sum_m m\, R(m) \left| e_I(m) \right|^2 }{\sum_m R(m) \left| e_I(m) \right|^2}
 \end{aligned}
 \label{eq:ref_direction_curved}
\end{equation} 
with $\sin(\chi_\text{ref}) = \langle m \rangle_\text{ref}/(nkR)$.
Equivalently to the planar case, the intensity transmission coefficient is $R(m) = \left| \rho_c(m) \right|^2$
The angular deflection due to the Fresnel filtering effect then is
\begin{equation}
\begin{aligned}
 \Delta \chi_\text{FF} &= \chi_\text{ref} - \chi_\text{in} \\
 &=  \arcsin\left( \frac{\langle m \rangle_\text{ref}}{nkR} \right) - \arcsin\left( \frac{m_0}{nkR} \right).
 \end{aligned}
 \label{eq:FF_curved}
\end{equation}

The Fresnel filtering effect in transmission can be deduced from the expectation value of the angular wavenumber in the reflected profile
\begin{equation}
 \langle m \rangle_\text{trans} = \frac{\sum_m m\, T(m) \left| e_I(m) \right|^2 }{\sum_m T(m) \left| e_I(m) \right|^2}
 \label{eq:trans_direction_curved}
\end{equation} 
with the intensity transmission coefficient $T(m)=1-R(m)$.
The angular shift $\Delta \chi^\text{trans}_\text{FF}$ is the difference between the mean angle of transmission calculated from the expectation value of the transmitted beam and the angle of transmission expected from Snell's law
\begin{equation}
 \begin{aligned}
  \Delta \chi^\text{trans}_\text{FF} &= \chi_\text{trans} - \arcsin(n\sin(\chi_\text{in} )) \\
  &= \arcsin\left( \frac{\langle m \rangle_\text{trans}}{kR} \right) - \arcsin\left( \frac{m_0}{kR} \right).
 \end{aligned}
 \label{eq:FF_curved_trans}
\end{equation}

Finally, the intensity of the reflected beam is
\begin{equation}
 \begin{aligned}
 \langle I \rangle_\text{ref} &= \frac{\sum_m e_R^*(m)\, e_R (m)}{\sum_m e_I^*(m)e_I(m)} \\
 &= \frac{\sum_m R(m) \left| e_I(m) \right|^2 }{\sum_m \left| e_I(m) \right|^2}.
 \end{aligned}
 \label{eq:I_ref_curved}
\end{equation} 

\subsection{The correction terms used for the asymmetrically deformed disks}
\label{app:curved}

For an evaluation of the formulas for the correction terms at curved interfaces, we chose
\begin{equation}
 e_I(m) = \frac{1}{\sqrt{2\pi}\sigma} \text{e}^{-\frac{(m-m_0)^2}{2\sigma^2}}
 \label{eq:gaussian_curved}
\end{equation}
as the transverse beam profile in angular momentum space which is a normal distribution in $m$ with mean value $m_0$ and standard deviation $\sigma$ similar to the Gaussian beam profile defined in Eq.~\eqref{eq:gaussian_planar} for planar interfaces.

\begin{figure}
 \includegraphics[width=\linewidth]{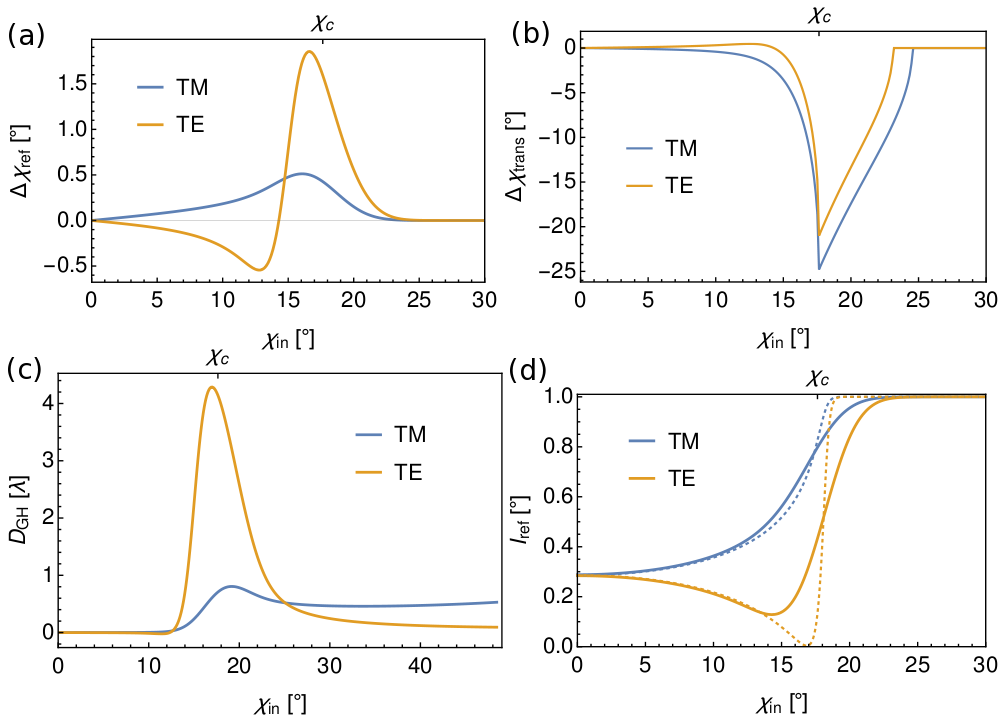}
 \caption{\label{fig:corrections_curved}
 Wave-inspired corrections at a curved interface with relative refractive index $n=3.3$ and size parameter $kR_0=147$ for both polarizations, TM (blue) and TE (orange), respectively, and for an angular momentum distribution according to Eq.~\eqref{eq:gaussian_curved} with standard deviation $\sigma=22$.
 }
\end{figure}

The resulting corrections calculated for a system with relative refractive index $n=3.3$ and size parameter $kR_0=147$ using the angular momentum distribution according to Eq.~\eqref{eq:gaussian_curved} with standard deviation $\sigma=22$ are shown in Fig.~\ref{fig:corrections_curved}.
These corrections are applied to the asymmetrically deformed disks discussed in Section \ref{sec:curved} in the main part of the paper.
\begin{acknowledgments}
 We thank the German Research Foundation (DFG) for funding within the Emmy-Noether programme. 
\end{acknowledgments}

\bibliography{Optical_microcavities.bib}

\end{document}